\documentstyle[aps,multicol,epsf]{revtex}

\begin{document}
\draft

\title{
Generic scale of the ``scale-free'' growing networks
}

\author{
S.N. Dorogovtsev$^{1, 2, \ast
}$, J.F.F. Mendes$^{1,\dagger}$, 
and A.N. Samukhin$^{2, \ddagger}$
}

\address{
$^{1}$ Departamento de F\'\i sica and Centro de F\'\i sica do Porto, Faculdade 
de Ci\^encias, 
Universidade do Porto\\
Rua do Campo Alegre 687, 4169-007 Porto, Portugal\\
$^{2}$ A.F. Ioffe Physico-Technical Institute, 194021 St. Petersburg, Russia 
}

\maketitle
   
\begin{abstract}
We show that the connectivity distributions $P(k,t)$ of scale-free growing 
networks ($t$ is the network size) have the generic scale -- the cut-off at 
$k_{cut} \sim t^\beta$.    
The scaling exponent $\beta$ is related to the exponent $\gamma$ of the connectivity distribution, $\beta=1/(\gamma-1)$. We propose the simplest model of scale-free growing networks and obtain the exact form of its connectivity distribution for any size of the network. We demonstrate that the trace of the initial conditions -- a hump at $k_h \sim k_{cut} \sim t^\beta$ -- may be found for any network size. 
We also show that there exists a natural boundary for the observation of the scale-free networks and explain why so few scale-free networks are observed in Nature.  
\end{abstract}


\begin{multicols}{2}
\narrowtext

%%%%%%%%%%%%%%%%%%%%%%%%%%%%%%%%%%%%%%%%%%%%%%%%%%%%%%%%%%%%%%%%%%%%%%%

A significant progress was made recently in the field of evolving networks 
\cite{r98,ajb99,fff99,krrt99,ha99,asbs00,bkm00}. 
It was observed that a number of growing networks in Nature (World-Wide Web, Internet, networks of scientific citations, collaboration nets, some networks in biology, etc.) 
are {\em scale-free}, i.e., 
their connectivity distribution is of a power-law form. Moreover, 
it was found that at least many of them {\em have to be scale-free}, otherwise growing networks are not resilient enough to random breakdowns \cite{ba00a,ceah00}. The infinite scale-free network with the connectivity distribution exponent $\gamma<3$ does not decay for {\em any} concentration (less than one) of randomly removed 
%nodes or 
links \cite{ceah00}. 

The proposed mechanism of 
%the 
self-organization of networks into 
%the 
scale-free structures, 
the 
preferential linking, is 
quite 
natural \cite{ba99}. 
New links of the growing networks are preferentially attached to nodes which already have many connections (connectivity $k$). In fact, it is the realization of a general principle -- {\em popularity is attractive}. Several types of preferential linking were proposed 
\cite{ba99,baj99,dm00,dms00,krl00,comm} which provide a variety of the $\gamma$ exponent values between $2$ and infinity. 

One should emphasize that only a few scale-free networks is known yet. The range of the values of the connectivity, in which the power-law behavior can be observed, is usually too narrow for a precise measurement of the exponent $\gamma$. It is unclear, why are so few scale-free networks observed? Why are the values of $\gamma$ for all of them only between $2$ and $3$? 
(Note that not any network has to be resilient, e.g., neither nodes nor links of collaboration networks are removable by definition \cite{n00}.)
In the present Letter, we answer these questions.

In previous papers, the connectivity distributions $P(k,t)$ of scale-free networks were calculated in the ``thermodynamic
%al 
limit'', i.e., in the limit of the large system size, $t$, which also plays the role of time, if one node is added at each increment of time. 
In this case, the distribution is stationary, and 
is of the form 
$P(k) \sim k^{-\gamma}$ in all range of large enough $k$, $k \gg 1$. Nevertheless, real networks are finite. 
The evolution of $P(k,t)$ to the stationary distribution turns to be non trivial.
We demonstrate below that, for finite networks, the power-law region of the connectivity distribution has the cut-off at $k_{cut} \sim t^\beta$, where $\beta=1/(\gamma-1)$.
We show that the trace of the initial conditions, i.e, of the initial configuration of the network -- the hump at $k_h \sim k_{cut} \sim t^\beta$ -- may be observed at any size of the network. 

This cut-off in the connectivity distribution allows 
%the 
observation of the power-law dependence only for very large networks. We show that for large values of $\gamma$, the power-law dependence is practically unobservable. 

Two answers have been already given to the question: {\em Why are the observed scale-free networks such as they are?} First, -- because their 
%growth proceeds with 
evolution is determined by 
the preferential linking 
mechanism \cite{ba99}. Second, -- because otherwise they would be unstable, weak against processes of decay and could not exist as united systems \cite{ba00a,ceah00}. Here, we propose the third answer, -- because {\em otherwise they would be unobservable}, i.e., it would be impossible to observe 
a 
%the 
power-law distribution.
%s. 

We demonstrate these features of the connectivity distribution using the simplest model of the scale-free growing network for which we present the exact solution -- the implicit form of the connectivity distribution for all sizes of the network. 
Also, we obtain them from general considerations. 
One should note that the introduced model is interesting by itself, so we present briefly main exact results for it. 

\begin{figure}
\epsfxsize=75mm
\epsffile{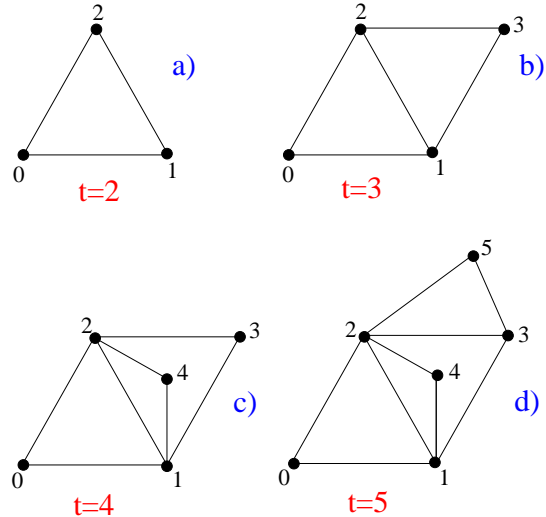}
\caption{
Illustration of the simplest model of scale-free growing networks. In the initial configuration, 
$t=2$, three sites are present, $s=0,1,2$ $(a)$. At each increment of time, a new node with two links is added. 
These links are attached to the ends of a randomly chosen link of the network.
}
\label{f1}
\end{figure}

Let us introduce the simplest model of the scale-free growing network with {\em undirected} links (see Fig. \ref{f1}). Initially ($t=2$), there are three nodes, $s=0,1,2$, each with the connectivity $2$. 
%, i.e. a ring. 
(The connectivity of a site is a number of its connections.)

(i) At each increment of time, a new node is added.

(ii) It is connected to both ends of a randomly chosen link by two undirected links.  

As far as we know, it is the simplest model of a scale-free network. 
The preferential linking arises in it not because of some special rule 
including a function of connectivity 
as in \cite{ba99}  
but naturally. Indeed, in the model that we consider, the probability that a node has the randomly chosen link attached to it is equal to the connectivity $k$ of the node divided by the total number of links, $2t-1$. Therefore, the evolution of the network is described by the following master equation,

\begin{equation}
\label{1}
p(k,s,t+1) = \frac{k-1}{2t-1} p(k-1,s,t) + 
%\left(1 - \frac{k}{2t-1}  \right) 
\frac{2t-1-k}{2t-1}
p(k,s,t)
,
\end{equation}
with the initial condition, $p(k,s=\{0,1,2\},t=2)=\delta_{k,2}$. Also, $p(k,t,t)=\delta_{k,2}$.
Here $p(k,s,t)$ is the probability that the site $0 \leq s \leq t$ has $k$ connections at time $t$. 
Note that this master equation and all the following ones are exact for all $t \geq 2$. Eq. (\ref{1}) has the form similar to that 
%one 
of the Barab\'{a}si-Albert's model \cite{ba99}. Therefore, 
one may expect that the scaling exponents of these models have to coincide.  

From Eq. (\ref{1}), we can obtain a number of usefull exact relations for our model.  
In particular, from Eq. (\ref{1}), one may find the equation for the average connectivity of an individual node, $\overline{k}(s,t)\equiv\sum_{k=2}^{t-s+2} k p(k,s,t)$: 

\begin{equation}
\label{2}
\overline{k}(s,t+1) = \frac{2t}{2t-1}\overline{k}(s,t) \ \ \ \ , \ \ \ \ \overline{k}(t,t)=2
\, .
\end{equation}
One can obtain easily its solution:

\begin{equation}
\label{3}
\overline{k}(s,t) = 2^{t-s+1} \frac{(t-1)!}{(s-1)!} \frac{(2s-3)!!}{(2t-3)!!} \,
\stackrel{s,t \gg 1}{\cong} \,
2\sqrt{\frac{t}{s}}
\, .
\end{equation}
Here, $s \geq 2$ and $\overline{k}(0,t)=\overline{k}(1,t)=\overline{k}(2,t)$. 
Hence, the scaling exponent $\beta$, 
defined through the relation, 
$\overline{k}(s,t) \propto (s/t)^{-\beta}$, equals $1/2$ like for the Barab\'{a}si-Albert's model. 

Also, one may find the average number $\overline{b}(s,s^\prime)$ of links between the sites $s$ and $s^\prime$ for any $s < s^\prime \leq t$, $0 < \overline{b}(s,s^\prime) \leq 1$. 
In fact, $\overline{b}(s,s^\prime)$ is the average of the element of the connectivity matrix over all possible realizations of the growth.   
The equation for this quantity is

\begin{equation}
\label{4}
\overline{b}(s,s^\prime+1) = \frac{1}{2t-1}\left[\sum_{u=0}^{s-1} \overline{b}(u,s) + 
\sum_{u=s+1}^{s^\prime} \overline{b}(s,u) \right] 
\, .
\end{equation}
Its exact solution for $s < s^\prime$ is of the form:

\begin{equation}
\label{5}
\overline{b}(s,s^\prime) = 2^{s^\prime-s} \frac{(s^\prime-2)!}{(s-1)!} \frac{(2s-3)!!}{(2s^\prime-3)!!} \,
\stackrel{s,s^\prime \gg 1}{\cong} \,
\frac{1}{\sqrt{s s^\prime}}
\, ,
\end{equation}
and $\overline{b}(0,s^\prime)=\overline{b}(1,s^\prime)=\overline{b}(2,s^\prime)=1$. 

We found exactly the connectivity distribution of the oldest nodes, 
$p(k,0,t)=p(k,1,t)=p(k,2,t)$:

\begin{equation}
\label{6}
p(k,2,t) = \frac{(k-1)}{2^{t-k}(t-k)!} \frac{(2t-k-2)!}{(2t-3)!!} \,
\stackrel{t \gg k}{\cong} \,
\frac{(k-1)}{2t}
\, .
\end{equation}
This relation turns to be useful for finding the total connectivity distribution. 
Also, one may obtain the relation, $p(2,s,t)=(2s-3)/(2t-3)$. The scaling form of $p(k,s,t)$ 
for $k,s,t \gg 1$ and $k\sqrt{s/t}$ fixed is obtained using the Z-transform for the connectivity, $k$. The scaling relation is of the form:

\begin{equation}
\label{7}
p(k,s,t) = \sqrt{\frac{s}{t}} \left( k\sqrt{\frac{s}{t}} \right) 
\exp\left(-k\sqrt{\frac{s}{t}} \right)
\, .
\end{equation}
This is a particular case of the corresponding scaling relations for the scale-free networks \cite{dms00}, 
see Eq.  (\ref{13}). 

%%%%%%
\end{multicols}
\widetext
\noindent\rule{20.5pc}{0.1mm}\rule{0.1mm}{1.5mm}\hfill
%%%%%%  

The matter of interest is the total connectivity distribution, 
$P(k,t)\equiv \sum_{s=0}^{t}p(k,s,t)/(t+1)$. The equation for it can be derived from Eq. (\ref{1}):

\begin{equation}
\label{8}
P(k,t) = \frac{t}{t+1}
\left[
\frac{k-1}{2t-3} P(k-1,t-1) + \left(1 - \frac{k}{2t-3}  \right) P(k,t-1)
\right]
+ \frac{1}{t+1} \delta_{k,2}
\, 
\end{equation}
with the initial condition $P(k,2)=\delta_{k,2}$. 

The exact solution of Eq. (\ref{8}) is:

\begin{equation}
\label{9}
P(k,t)  =  \frac{24}{k(k+1)(k+2)}\,\frac{1}{(t+1)(2t-3)!!}\, \frac{(2t-k-2)!}{2^{t-k}(t-k)!} 
\left\{ 
(t-k)\left[t + \frac{(k-2)(k+1)}{4} \right] 
+ \frac{(k-1)k(k+1)(k+2)}{8}  \right\}
\, .
\end{equation}
%%%
\hfill\rule[-1.5mm]{0.1mm}{1.5mm}\rule{20.5pc}{0.1mm}
\begin{multicols}{2}
\narrowtext
%%%
\noindent
One may check Eq. (\ref{9}) inserting it directly into Eq. (\ref{8}). We obtained Eq. (\ref{9}) using the distribution function $\tilde{P}(k,t) \equiv \sum_{s=3}^{t}p(k,s,t)/(t-2)$, which looks less cumbersome than $P(k,t)$ and may be found without great problems, and the expression for $p(k,2,t)$, Eq. (\ref{6}). 

From Eq. (\ref{8}) with $t \to \infty$, it follows the equation for the stationary distribution, $P(k)$, 

\begin{equation}
\label{10}
(k-1) P(k-1) - (k+2) P(k) + 2\delta_{k,2} = 0
\, 
\end{equation}
where the solution is:

\begin{equation}
\label{11}
P(k) = \frac{12}{k(k+1)(k+2)}
\, .
\end{equation}
Eq. (\ref{11}) is similar to the form of the stationary connectivity distribution found for  
the Barab\'{a}si-Albert's model \cite{dms00,krl00}. One sees that $\gamma=3$.

\begin{figure}
\epsfxsize=75mm
\epsffile{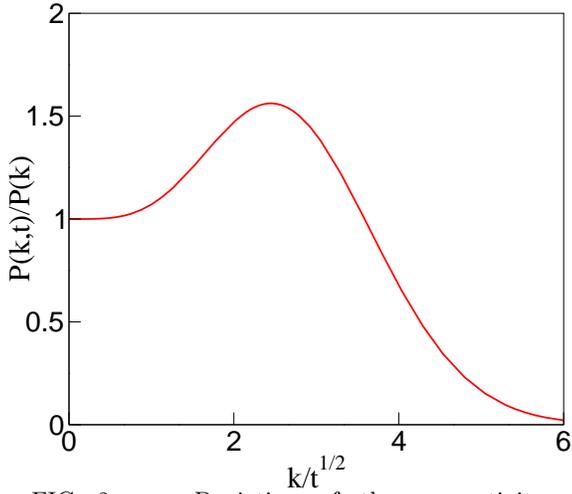}
\caption{
Deviation of the connectivity distribution of the finite-size network from the stationary one, 
$P(k,t)/P(k,t \to \infty)$, vs $k/\sqrt t$. The form of the hump depends on the initial configuration.
}
\label{f2}
\end{figure}

Our aim is to find how the stationary distribution is reached. 
From Eq. (\ref{9}), for $t \gg k \gg 1$, one gets

\begin{equation}
\label{12}
P(k,t) = P(k) \left[ 1 + \frac{1}{4}\frac{k^2}{t} + \frac{1}{8}\left(\frac{k^2}{t}\right)^2  \right] \exp\left\{ -\frac{1}{4}\frac{k^2}{t} \right\}
\, .
\end{equation}
The factor $P(k,t)/P(k) \equiv g(k/\sqrt t)$ depends only on the combination $k/\sqrt t$. 
Therefore, the peculiarities of the distribution induced by the size effects never disappear but only move with increasing time in the direction of large connectivity.
The function $g(k/\sqrt t)$ is close to $1$ for $k < \sqrt t$, has a 
%broad 
hump at $k_{max}$ between $\sqrt t$ and $4\sqrt t$ with a maximum at $k_{max}/\sqrt t = \sqrt 6 = 2.449\ldots$, 
$g(k_{max}/\sqrt t)=7\, e^{-3/2}=1.562\ldots$, and the cut-off at $k \sim 4\sqrt t$ (see Fig. \ref{f2}). Hence, the power-law behavior 
is observable only in a rather narrow region, $1 \ll k \ll \sqrt t$.  

One may check that the form of the hump in Fig. \ref{f2} depends on the initial conditions. 
In our case, the evolution starts from the configuration shown in  Fig. \ref{f1},a.
If we start the growth from another configuration, the form would be different.   

We have demonstrated above the size-dependence of the connectivity distribution using the exactly solvable example. What are the general reasons of such behavior of the scale-free networks? 
Let us obtain the general estimation of the distribution cut-off position for an arbitrary scale-free network.  

Measuring of connectivity distributions is always impeded by 
the 
strong fluctuations at large $k$. 
The reason of such fluctuations is 
the 
poor statistics in this region. One can easily estimate 
the characteristic value, $k_f$, above which the fluctuations are strong.  
Let $n$ be the total number of links of the network, and $\gamma>2$. For the linearly growing network, $n=mt$, where $m$ is the number of links added at each increment of time. If $P(k) \sim k^{-\gamma}$, 
$n k_f^{-\gamma} \sim 1$. Therefore, $k_f \sim n^{1/\gamma}$. One may improve the situation using the cumulative distributions, $
%P_{cum}(k) 
\equiv \int_k^\infty dk P(k)$, instead of $P(k)$. 
Also, in simulations, one may make a lot of runs to increase the statistics. 
Nevertheless, one can not pass the cut-off, $k_{cut}$, that we discuss. 
This cut-off is the real barrier for the observation of the power-law dependence.

We have shown that the connectivity distribution of individual sites is an exponentially decreasing function at large $k$ (see Eq. (\ref{7})). For the scale-free networks, it 
can be written in the general scaling form \cite{dms00}: 

\begin{equation}
\label{13}
p(k,s,t) = \left(\frac{s}{t}\right)^\beta  
f\left(k \left(\frac{s}{t}\right)^\beta \right)
\, ,
\end{equation}
where $f(x)$ is a scaling function, and 
the relation \cite{dm00,dms00} between the exponents $\beta$ and $\gamma$ is 

\begin{equation}
\label{14}
\beta(\gamma-1)=1  
\, .
\end{equation} 
In the particular case of the proposed model, $f(x) = x \exp(-x)$. 
The exponent $\beta$ also figures in the relation for the average connectivity, 
$\overline{k}(s,t) \propto (s/t)^{-\beta}$.  
It follows from Eq. (\ref{13}) that the cut-off of the total distribution is determined by the connectivity distribution of the individual nodes with the smallest number $s$, i.e., by the oldest ones. Therefore, $k_{cut}(1/t)^\beta \sim const$ and $k_{cut} \sim t^\beta=t^{1/(\gamma-1)}$. 
For the considered model, $\beta=1/2$, see Eq. (\ref{7}).
The connectivity distributions of the oldest nodes (and the quantity of them) depend strongly on the initial conditions. Hence, the part of the total connectivity distribution near the cut-off depends strongly on this factor. 
Now it becomes obvious why there are no scale-free networks with large values of $\gamma$. 
Indeed, the power-law dependence of the distribution can be observed only if it exists for at least $2$ or $3$ decades of the connectivity. For this, the networks have to be large, 
$t > 10^{2.5(\gamma-1)}$. But there is only a few large networks in Nature! If $\gamma>3$, one practically has no chances to find the scale-free behavior. 

\begin{figure}
\epsfxsize=85mm
\epsffile{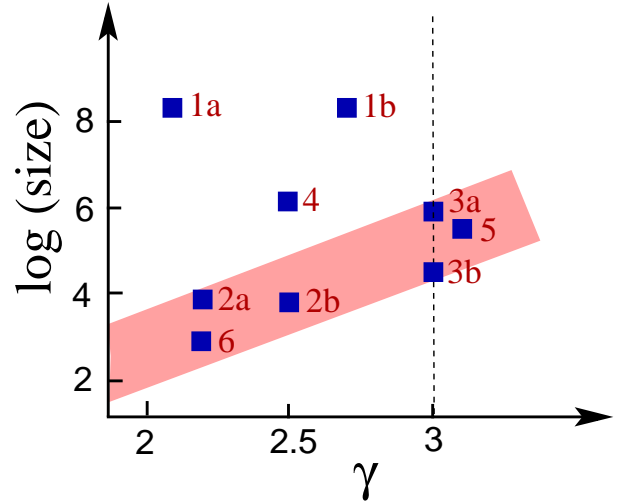}
\caption{
Log-linear plot of the size vs the $\gamma$ exponent value of the all observed scale-free networks.
The line $\log_{10} t \sim 2.5(\gamma-1)$ is the finite-size boundary for the observation of the power-law 
connectivity distributions. The dashed line, $\gamma=3$, is the resilience boundary. 
This boundary is 
important for those growing networks which have to be stable to random breakdowns. 
The points: 1$a$ and 1$b$ are obtained for incoming- and outgoing links of the pages 
of the World-Wide Web \protect\cite{krrt99,bkm00} (also, $\gamma_{in}=2.1$ and $\gamma_{out}=2.45$ 
were obtained from the complete map of the nd.edu domain of the Web, 
$325,729$ nodes  \protect\cite{ba99}, 
$\gamma_{in}=1.94$ was obtained for the domain level of the Web in spring 1997 
\protect\cite{ah00}), 
2$a$ is for outgoing links for the inter-domain structure of the Internet and  
2$b$ is for outgoing links for the Internet at the router level \protect\cite{fff99}, 
3$a$ and 3$b$ are for citations of the ISI data base and Phys. Rev. D \protect\cite{r98}, 
4 is for the collaboration network of MEDLINE \protect\cite{n00}, 5 is for the 
collaboration network of movie actors \protect\cite{ab00b}, (also, $\gamma=2.3$ was 
obtained for this network in \protect\cite{asbs00}) 6 is for incoming and outgoing links 
of the networks of the metabolic reactions \protect\cite{privat}. The precision of the 
upper points is about $\pm 0.05$ and is much worse for points in the dashed region.
}
\label{f3}
\end{figure}

In Fig. \ref{f3}, in a log-linear scale, we present  
the sizes of all known scale-free networks vs their $\gamma$ exponent values. 
The plotted points are inside of the region restricted by the lines: $\gamma=2$, $\log_{10} t \sim 2.5(\gamma-1)$, and by 
%in the triangle, one of the side of which is the line $\gamma=2$, 
%the other -- is $\log_{10} t \sim 2.5(\gamma-1)$, and the third one -- is 
the logarithm of 
the size of the largest scale-free network in Nature -- the World-Wide Web, -- $\log_{10} t \sim 9$. 
%Note that practically the all scale-free networks produced 
%by the preferential linking have $\gamma>2$.

We have demonstrated that the form of the connectivity distribution is influenced by initial conditions even for large networks. Therefore, it is hard to obtain the values of the scaling exponents with high precision both from experimental data and simulations. 
One should note that including the aging of nodes, breaking of links, or disappearing of nodes suppresses the effect of the initial conditions and removes the hump (see the plots of the connectivity distributions in  \cite{dm00}).

In conclusion, we have described the size effect on the connectivity distribution of the scale-free growing networks. We have shown that the scale-free networks have 
%the 
a 
generic scale -- 
the size-dependent cut-off, $k_{cut} \sim t^{1/(\gamma-1)}$ of the connectivity distribution. 
This cut-off impedes observations of the power-law dependence even for large networks. 
For large $\gamma$, such observations are impossible. 
If $\gamma \to 2$, then $k_{cut} \sim t$, so, in fact, the cut-off is absent.
We have estimated the region of the network sizes and the values of the exponent $\gamma$ in which the power law is visible. All found scale-free networks are in this region. We have shown that the trace of the initial configuration of the network -- the hump near $k_{cut}$ -- may be observed for all sizes of the network. We have demonstrated such behavior using the simplest model of 
%the 
a scale-free growing network. It turned to be possible to find the exact solution of it for any size of the network. Also, these results have been obtained from 
%the 
general considerations. 
In fact, they are general and applicable to systems displaying power-law distributions.  

The proposed model belongs to the class of the exactly solvable scale-free growing 
networks. One can consider another simple model. 
Instead of the connecting of a new node with the ends of a randomly chosen link of the network, one may connect it each time with all three vertex nodes of a randomly chosen 
triangle of links. (Note that we forbid multiple links.) Such a model has the same scaling exponents as the considered one.

It follows from our results, that one can not see the scale-free networks with large $\gamma$. 
Also, if we do not observe the scale-free connectivity distributions of some growing network, 
this does not mean at all that it is not a scale-free one. There is a chance that the power-law behavior will be found after some time, when the network will grow up.  
\\

SND thanks PRAXIS XXI (Portugal) for a research grant PRAXIS XXI/BCC/16418/98. JFFM 
was partially supported by the project PRAXIS/2/2.1/FIS/299/94. We also thank A.-L. Barab\'{a}si for providing us with the preprint \cite{privat} before publication. 
We are grateful to S. Bornholdt and H. Ebel for a useful remark on the measured values of the $\gamma$ exponent for the World-Wide Web.
\\

\noindent
$^{\ast}$      E-mail address: sdorogov@fc.up.pt \\
$^{\dagger}$   E-mail address: jfmendes@fc.up.pt \\
$^{\ddagger}$  E-mail address: alnis@samaln.ioffe.rssi.ru

\newpage

%\newpage

\end{multicols}


\begin{references}

\bibitem{r98}  S. Redner, Eur. Phys. J. B {\bf 4}, 131 (1998).

\bibitem{ajb99}  R. Albert, H. Jeong, and A.-L. Barab\'{a}si, Nature {\bf 401}, 
130 (1999)

\bibitem{fff99} M. Faloutsos, P. Faloutsos, and C. Faloutsos, Comput. Commun. Rev. {\bf 29}, 251 (1999). 

\bibitem{krrt99} R. Kumar, P. Raghavan, S. Rajagopalan, and A. Tomkins, Proc. of the 25th VLDB Conference (Edinburgh, 1999), 639-650.

\bibitem{ha99} B.A. Huberman and L.A. Adamic, Nature {\bf 401}, 131 (1999).

\bibitem{asbs00}  L.A.N. Amaral, A. Scala, M. Barthelemy, and H.E. Stanley,
cond-mat/0001458, to appear in Proc. Nat. Acad. Sci. (USA). 

\bibitem{bkm00} A. Broder, R. Kumar, F. Maghoul, P. Raghavan, S. Rajagopalan, R. Stata, A. Tomkins, and J. Wiener, Proc. of the 9th WWW Conference (Amsterdam, 2000), 309.  
 
\bibitem{ba00a} R. Albert, H. Jeong, and A.-L. Barab\'{a}si, Nature {\bf 406}, 
378 (2000).

\bibitem{ceah00} R. Cohen, K. Erez, D. ben-Avraham, and S. Havlin, cond-mat/0007048.

\bibitem{ba99}  A.-L. Barab\'{a}si and R. Albert, Science {\bf 286}, 509
(1999). 

\bibitem{baj99} A.-L. Barab\'{a}si, R. Albert, and H. Jeong, Physica A {\bf 272}, 173
(1999).

\bibitem{dm00}  S.N. Dorogovtsev and J.F.F. Mendes, 
%cond-mat/\\0001419, 
Phys. Rev. E {\bf 62}, 1842 (2000); 
%cond-mat/0005050, 
Europhys. Lett. {\bf 52}, 33 (2000).

\bibitem{dms00} S.N. Dorogovtsev, J.F.F. Mendes, and A.N. Samukhin,  
cond-mat/0004434.

\bibitem{krl00} P.L. Krapivsky, S. Redner, and F. Leyvraz, cond-mat/0005139, 
to appear in Phys. Rev. Lett.

\bibitem{comm} S. Bornholdt and H. Ebel, cond-mat/0008465; S.N. Dorogovtsev, J.F.F. Mendes, and A.N. Samukhin, cond-mat/0009090.

\bibitem{ab00b}  R. Albert and A.-L. Barab\'{a}si, cond-mat/0005085.

\bibitem{n00} M.E.J. Newman, cond-mat/0007214.

\bibitem{ah00} L.A. Adamic and B.A. Huberman, Science {\bf 287}, 2115a (2000).

\bibitem{privat} H. Jeong, B. Tombor, R. Albert, Z. N. Oltvai 
and A-L. Barab\'{a}si, Nature, in press.


%\bibitem{hppj98}  B.A. Huberman, P.L.T. Pirolli, J.E. Pitkow and R.J.
%Lukose, Science {\bf 280}, 95 (1998).

%\bibitem{ls98}  J. Lahererre and D. Sornette, Eur. Phys. J. B {\bf 2}, 525
%(1998).

%\bibitem{mn00} C. Moore and M.E.J. Newman, Phys. Rev. E {\bf 61}, 5678 (2000). 
 
%\bibitem{b1} P. Baran, {\it Introduction to Distributed Communications Networks}, 
%RM-3420-PR, August 1964, $<$http://www.rand.org/publications/RM/baran.list.html$>$.

%\bibitem{baj99} A.-L. Barab\'{a}si, R. Albert, and H. Jeong, Physica A {\bf 272}, 173
%(1999).

%\bibitem{dm00b} S.N. Dorogovtsev and J.F.F. Mendes, 
%cond-mat/\\0005050, 
%Europhys. Lett. (2000), to be published. 

 

\end{references}
\end{document}